\documentclass[12pt]{article}

\def\hybrid{\topmargin 0pt      \oddsidemargin 0pt
	\headheight 0pt \headsep 0pt
	\textwidth 6.25in       % A4 paper
        \textheight 9.5in       % A4 paper
	\marginparwidth .875in
	\parskip 5pt plus 1pt   \jot = 1.5ex}

\catcode`\@=11
\def\marginnote#1{}

\newcount\hour
\newcount\minute
\newtoks\amorpm
\hour=\time\divide\hour by60
\minute=\time{\multiply\hour by60 \global\advance\minute by-\hour}
\edef\standardtime{{\ifnum\hour<12 \global\amorpm={am}%
	\else\global\amorpm={pm}\advance\hour by-12 \fi
	\ifnum\hour=0 \hour=12 \fi
	\number\hour:\ifnum\minute<10 0\fi\number\minute\the\amorpm}}
\edef\militarytime{\number\hour:\ifnum\minute<10 0\fi\number\minute}

\def\draftlabel#1{{\@bsphack\if@filesw {\let\thepage\relax
   \xdef\@gtempa{\write\@auxout{\string
      \newlabel{#1}{{\@currentlabel}{\thepage}}}}}\@gtempa
   \if@nobreak \ifvmode\nobreak\fi\fi\fi\@esphack}
	\gdef\@eqnlabel{#1}}
\def\@eqnlabel{}
\def\@vacuum{}
\def\draftmarginnote#1{\marginpar{\raggedright\scriptsize\tt#1}}

\def\draft{\oddsidemargin -.5truein
	\def\@oddfoot{\sl preliminary draft \hfil
	\rm\thepage\hfil\sl\today\quad\militarytime}
	\let\@evenfoot\@oddfoot \overfullrule 3pt
	\let\label=\draftlabel
	\let\marginnote=\draftmarginnote
   \def\@eqnnum{(\theequation)\rlap{\kern\marginparsep\tt\@eqnlabel}%
\global\let\@eqnlabel\@vacuum}  }

\def\numberbysection{\@addtoreset{equation}{section}
	\def\theequation{\thesection.\arabic{equation}}}

\usepackage{epsfig}
\catcode`@=12
\relax

\def\beq{\begin{equation}}
\def\eeq{\end{equation}}
\def\bea{\begin{eqnarray}}
\def\eea{\end{eqnarray}}

\relax
\hybrid
\begin{document}
\begin{titlepage}
\begin{center}
October~2001 \hfill    PAR--LPTHE 01/?? \\[.5in]
{\large\bf Coupled Models $WD^{(p)}_{3}$. Their Fixed Points }\\[.3in] 

        {\bf Vladimir S. Dotsenko, Xuan Son Nguyen 
             and Raoul Santachiara } \\ [.3in] 

       {\it LPTHE\/}\footnote{Unit\'e Mixte de Recherche CNRS
UMR 7589.},\\
        {\it  Universit{\'e} Pierre et Marie Curie, PARIS VI\\
              Universit{\'e} Denis Diderot, PARIS VII\\
	      Bo\^{\i}te 126, Tour 16, 1$^{\it er}$ {\'e}tage \\
	      4 place Jussieu,
	      F-75252 Paris CEDEX 05, FRANCE}\\

\end{center}
\vskip .15in
\centerline{\bf ABSTRACT}
\begin{quotation}

{\small $N$ conformal theory models $WD^{(p)}_{3}$ coupled locally
by their energy operators are analyzed by means of a perturbative
renormalization group. New non-trivial fixed points are found.}

\end{quotation}
\end{titlepage}

\newpage

In the continuum limit the model $WD_{n}$ could be defined as:
\beq
A=\int d^{2}x\{\sum^{n}_{a=1}\partial\vec{\varphi}\bar{\partial}\vec
{\varphi}+\sum^{n}_{a=1}(e^{i\vec{\alpha}^{+}_{a}\vec{\alpha}}+
e^{i\vec{\alpha}^{-}_{a}\vec{\varphi}})\}
\eeq
\beq
\vec{\varphi}=\{\varphi_{1}, \varphi_{2},...,\varphi_{n}\},
n=3,4,5,...
\eeq
$\{\vec{\alpha}^{\pm}_{a}\}$ are simple roots of the algebra $D_{n}$.

In this class of models find themselves:

-- Toda models

-- Loop models, still to be found.

Corresponding conformal theory, which is based on $WD_{n}$, chiral
algebra, has been found by V.Fateev and S.Lukyanov [1]. The theory
contains minimal models:
\beq
M_{p}\sim WD^{(p)}_{n}
\eeq
with the central charge:
\bea
c_{p}=n(1-\frac{(2n-1)(2n-2)}{p(p+1)}),\\ p=2n-1, 2n, 2n+1,...\infty
\eea

The most simple model in this class is $WD_{3}$,
\beq
\vec{\varphi}=\{\varphi_{1},\varphi_{2},\varphi_{3}\}
\eeq

-- Coulomb gas of a 3-component field $\vec{\varphi}$;

-- Minimal models:
\beq
WD^{(p)}_{3}, p=5,6,7,...\infty
\eeq
with the geometry of screening operators in (1) given in Fig.1.

\begin{figure}[ht]
\epsfig{file=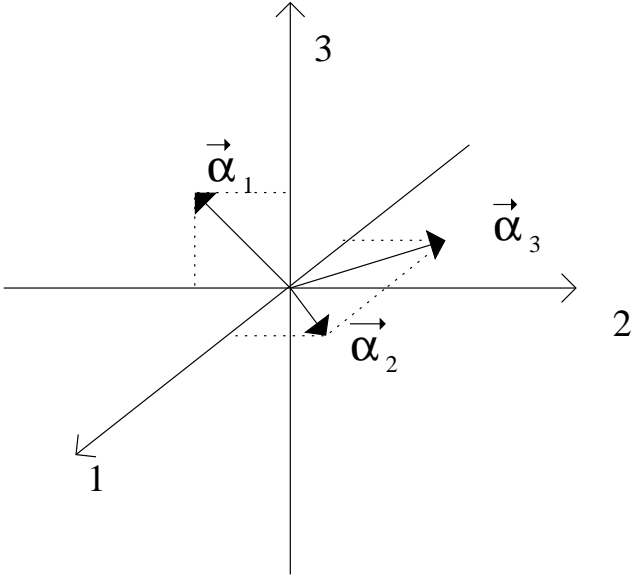} 
\caption{Screening operators geometry for $WD^{(p)}_{3}$}
\end{figure}

The model which is close, in a sense, to $WD_{3}$ is the loop model
of Jacobsen and Kondev [2]. Its 3-component Coulomb gas has the
geometry of screenings given in Fig.2.

\begin{figure}[ht]
\epsfig{file=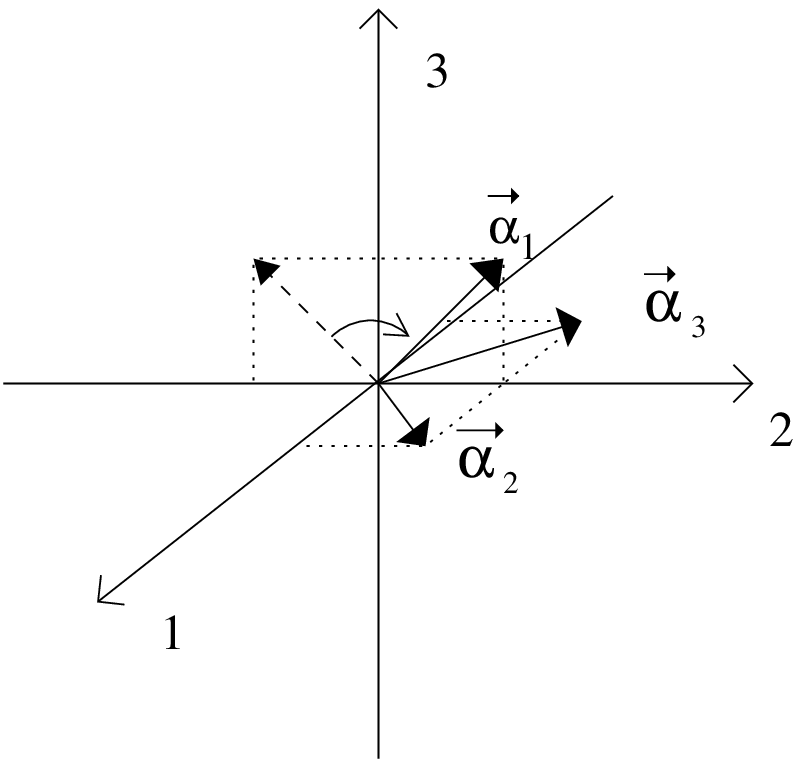} 
\caption{Screening operators geometry for the loop model of Jacobsen and Kondev }
\end{figure}

\underline{The energy operator of $WD^{(p)}_{3}$.}

The model contains the degenerate representations, the operators:
\beq
V_{\vec{\beta}}=e^{i\vec{\beta}\vec{\varphi}}
\eeq
with
\beq
\vec{\beta}=\vec{\beta}_{(n'_{1},n_{1})(n'_{2},n_{2})(n'_{3}n_{3})}=
\sum^{3}_{a=1}(\frac{1-n'_{a}}{2}\alpha_{-}+\frac{1-n_{a}}{2}\alpha_{+})
\vec{\omega}_{a}
\eeq
\beq
\alpha_{\pm}:\quad
\vec{\alpha}^{\pm}_{a}=\alpha_{\pm}\vec{e}_{a},\quad
|\vec{e}_{a}|=1
\eeq
For $WD^{(p)}_{3}$,
\beq
\alpha^{2}_{+}=(\alpha_{-})^{-2}=\frac{1+p}{p}
\eeq
\beq
\vec{\omega}_{a}:\quad \vec{e}_{a}\vec{\omega}_{b}=\delta_{ab}
\eeq
The conformal dimensions of these operators are given by the formula:
\beq
\Delta_{(n'_{1}n_{1})(n'_{2}n_{2})(n'_{3}n_{3})}=\vec{\beta}^{2}_{(...)}
-2\vec{\alpha}_{0}\vec{\beta}_{(...)}
\eeq
where
\beq
2\vec{\alpha}_{0}=(\alpha_{+}+\alpha_{-})\sum^{3}_{a=1}\vec{\omega}_{a}
\eeq
With this formula one finds that the most relevant operator in the neutral
sector (the operators (8) with $\vec{\beta}$ belonging to the plane
2,3 in Fig.1) corresponds to $\vec{\beta}_{(2.1)(1.1)(1.1)}$. It is 
natural to identify it with an energy operator of the model:
\beq
\varepsilon\sim V_{\vec{\beta}},\quad\vec{\beta}=\vec{\beta}_{(2.1)
(1.1)(1.1)}
\eeq

One finds:
\beq
\Delta_{\varepsilon}=\Delta_{(2.1)(1.1)(1.1)}=\frac{5}{2}\alpha^{2}_{-}
-2
\eeq
With $\alpha^{2}_{-}=p/(p+1)$, eq.(11),
\beq
\Delta_{\varepsilon}=\frac{1}{2}-\frac{5}{2}\epsilon
\eeq
where we have defined a perameter:
\beq
\epsilon=\frac{1}{p+1}
\eeq
The physical dimension of the operator $\varepsilon$ (conformal
dimension doubled) is:
\beq
\Delta_{\varepsilon}^{ph} = 1-5\epsilon
\eeq

\underline{$N$ coupled models $WD^{(p)}_{3},\quad p\gg 1$.}

\beq
A=\sum^{N}_{1}A_{0}+g\int d^{2}x\sum^{N}_{a\neq b}\varepsilon_{a}(x)
\varepsilon_{b}(x)
\eeq
-- the interaction term is slightly relevant, in the case of $p\gg 1,
\quad \epsilon\ll 1,$ eq.(19). The theory is accessible for the
analysis by a perturbative RG. One find a phase diagramm which is
fairly interesting.

Important observation:
\bea
\varepsilon(x)\varepsilon(x')&=&\frac{1}{|x-x'|^{2\Delta_{\varepsilon}}}+
\frac{D^{\Phi}_{\varepsilon\varepsilon}}{|x-x'|^{2\Delta_{\varepsilon}-
\Delta_{\Phi}}}\Phi(x')+... \\
\Delta_{\Phi}&=&2-B\epsilon
\eea
$B$ is a numerical constant. This operator algebra implies that another 
slightly relevant perturbation will be generated by the RG. The initial 
action, eq.(20), is unstable w.r.t. RG evolution. The generalized action,
which is stable (renormalizable), is:
\beq
A=\sum^{N}_{1}A_{0}+g\int d^{2}x\sum^{N}_{a\neq b}\varepsilon_{a}(x)
\varepsilon_{b}(x)+\lambda\int d^{2}x\sum^{N}_{a=1}\Phi_{a}(x)
\eeq

\underline{Model with disorder, $N\rightarrow 0$.}

One finds the following RG flow: Initial conditions $\{g_{0}\neq 0,
\lambda_{0}=0\} \Rightarrow $ Fixed point $\{ g^{\ast}=0, \lambda^{\ast}
\neq 0\}$, Fiq.3.
\begin{figure}[ht]
\epsfig{file=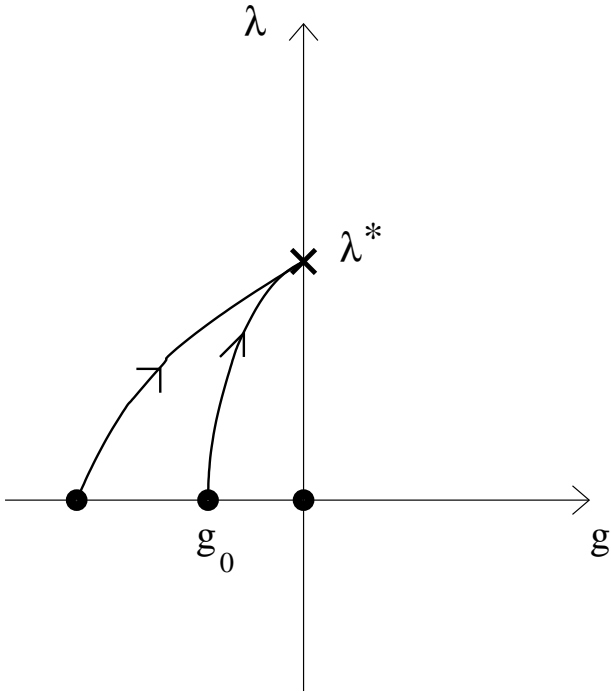} 
\caption{RG flow for the model with disorder}
\end{figure}

Checking the value of the central charge at the new fixed
point one finds that this flow corresponds to:
\beq
WD^{(p)}_{3}\mbox{with disorder} \Rightarrow WD^{(p-1)}_{3}\mbox
{without disorder}
\eeq
The details could be found in the paper [3].

\underline{Coupled models, $N\geq 2$.}

With the RG equations for the generalized action in eq.(23) one finds
the flows shown in Fig.4 [3]. There are 3 fixed points, in addition to the 
trivial one $g=\lambda=0$:
\begin{figure}[ht]
\epsfig{file=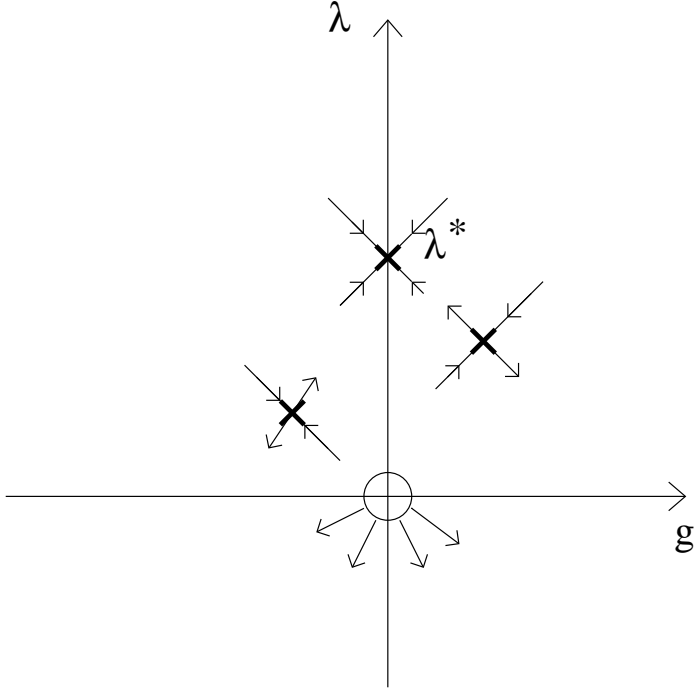} 
\caption{RG flows for the $N$ coupled models}
\end{figure}
1. $\lambda^{\ast}\neq 0,\quad g^{\ast}=0$

2. $\lambda^{\ast}_{1},\quad g^{\ast}_{1}\neq 0$

3. $\lambda^{\ast}_{2},\quad g^{\ast}_{2}\neq 0$

Point 1: $N$ decoupled models $WD^{(p-1)}_{3}$. 

Pont 2 and 3: the models are coupled. 
The symmetry of the critical theories at these points is $S_{3}$.
The corresponding conformal theory is unknown.

\underline{Passing in the vicinity of the point 3.}

When the initial values of the couplings are chosen appropriately, the
RG flow  makes first the couplings to approach the point 3, Fig.5, and
then, if the point is slightly missed to the right, the flow takes 
finally the couplings to the point 1.
\begin{figure}[ht]
\epsfig{file=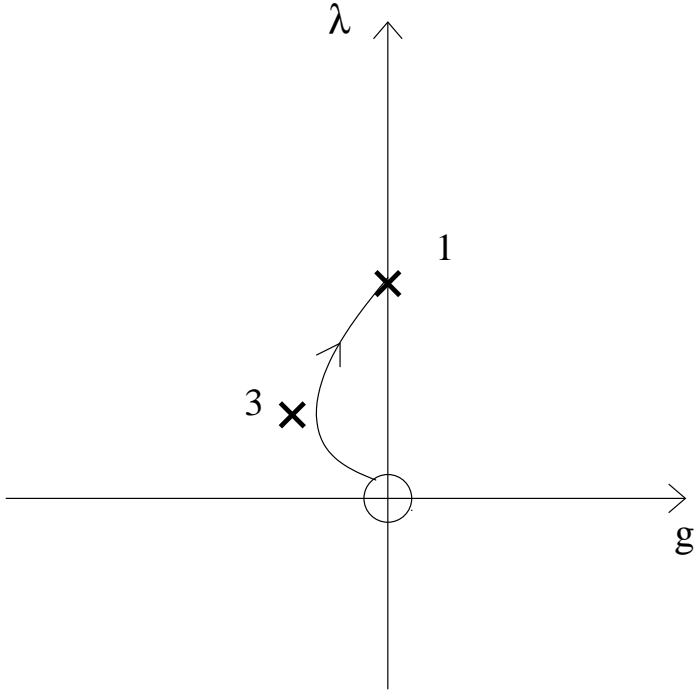} 
\caption{RG flow in the vicinity of the new fixed point}
\end{figure}
\begin{figure}[ht]
\epsfig{file=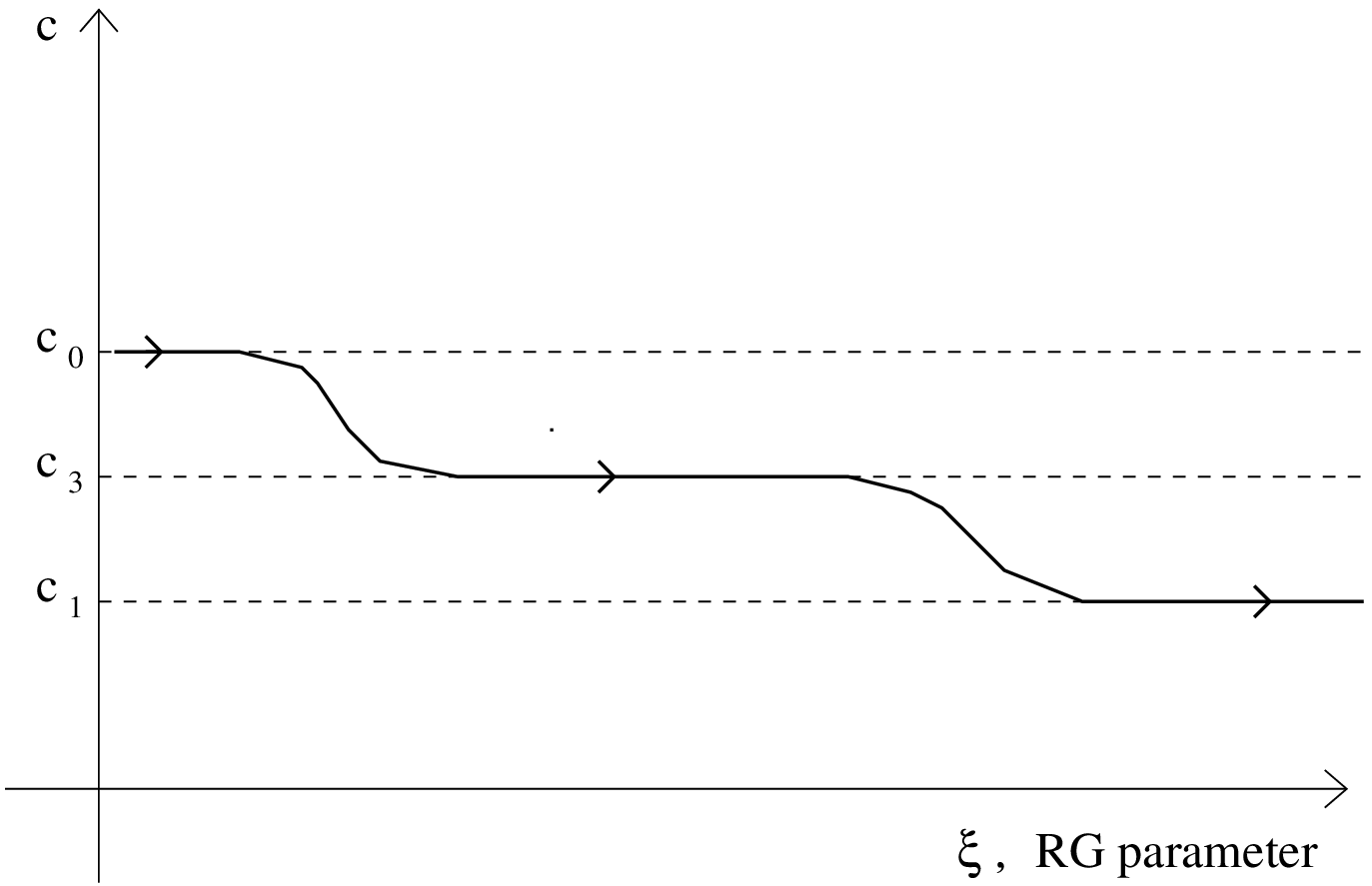} 
\caption{Evolution of the central charge along the RG flow}
\end{figure}
 The evolution of the effective
central charge ($C$ function of A.Zamolodchikov) along this flow is
shown in Fig.6. One observes three plateaux, signaling presence of
three fixed points which are being visited successively, Fig.6.
 This behaviour is reminiscent of the starecase model, studied by Al.B
Zamolodchikov [4].


\newpage
\small

\begin{thebibliography}{10}


  \bibitem{lf}   S.L.Lukyanov and V.A.Fateev, 
                       Sov. Sci. Rev. A Phys. {\bf 15} (1990) 1-117.

  \bibitem{jk}   J.~L.~Jacobsen and J.~Kondev,
                       Nucl.~Phys.~B {\bf 532} [FS], 635 (1998).

  \bibitem{dns}        Vl.~S.~Dotsenko, X.~S.~Nguyen and R.~Santachiara,
                       hep-th/0104197, Nucl.Phys.B, in print.

 \bibitem{AlbZama}
                 Resonance~Factorized~Scattering~and~Roaming~Trajectories,~ENS-
                 LPS~ 335 (1991)

\end{thebibliography}
\end{document}